\documentclass[preprint,review,12pt]{./elsarticle}
\usepackage{amssymb}
\usepackage{amsmath}
\usepackage{graphicx}
\usepackage{dcolumn}
\usepackage{bm}
\usepackage{epsfig}
\usepackage[T1]{fontenc}
\usepackage{ae,aecompl}
\usepackage{appendix}
\usepackage{color}

\journal{Physics Letters A}
\begin{document}
\begin{frontmatter}

\title{FFT-LB modeling of thermal liquid-vapor systems}

\author[a,b,c]{Yanbiao Gan}
\author[b]{Aiguo Xu \footnote{Corresponding author. Email address:
Xu\_Aiguo@iapcm.ac.cn}}
\author[b]{Guangcai Zhang}
\author[a]{Yingjun Li \footnote{Corresponding author. Email address:
lyj@aphy.iphy.ac.cn}}
\address[a]{ State Key Laboratory for
GeoMechanics and Deep Underground Engineering, SMCE, China
University of Mining and Technology (Beijing), Beijing 100083,
P.R.China \\}
\address[b]{National Key Laboratory of Computational Physics, \\
Institute of Applied Physics and Computational Mathematics, P. O.
Box 8009-26, Beijing 100088, P.R.China \\}
\address[c]{North China Institute of Aerospace Engineering,
Langfang 065000, P. R. China}

\begin{abstract}
We further develop a thermal LB model for multiphase flows. In the
improved model, we propose to use the windowed FFT and its inverse
to calculate both the convection term and external force term. By
using the new scheme, Gibbs oscillations can be damped effectively
in unsmooth regions while the high resolution feature of the
spectral method can be retained in smooth regions. As a result,
spatiotemporal discretization errors are decreased dramatically and
the conservation of total energy is much better preserved. A direct
consequence of the improvements is that the unphysical spurious
velocities at the interfacial regions can be damped to neglectable
scale. With the new model, the phase diagram of the liquid-vapor
system obtained from simulation is more consistent with that from
theoretical calculation. Very sharp interfaces can be achieved. The
accuracy of simulation results is also verified by the Laplace law.
The high resolution, together with the low complexity of the FFT,
endows the proposed method with considerable potential, for studying
a wide class of problems in the field of multiphase flows and for
solving other partial differential equations.
\end{abstract}

\end{frontmatter}

\section{Introduction}

During the past two decades, the lattice Boltzmann (LB) method has
been developed rapidly and has been successfully applied to various
fields \cite{Succi-Book}, ranging from magnetohydrodynamics \cite
{Chensy-PRL-1991,Succi-PRA-1991,CICP-2008}, to flows of suspensions
\cite {suspension-1,suspension-2}, flows through porous media \cite
{Succi-epl-1989-porous media,porous media-pre-2002}, compressible
fluid dynamics \cite{Watari,Xu-compressible,QuKun,LiQing-PLA-2009,
Xu-compressible-2,Xu-compressible-3}, wave propagation \cite%
{YanGW-JCP-2000-WAVE,YanGW-PRE-wave}, etc. Apart from the fields
listed above, this versatile method is particularly promising in the
area of multiphase systems \cite{chromodynamic model,
shanchen,Yeomans, HCZ model,XGL1,Succi-PRE-2007,Sofonea-multiphase}.
This is partly owing to its intrinsic kinetic nature, which makes
the inter-particle interactions (IPI) be incorporated easily and
flexibly, and, in fact, the IPI is the underlying microscopic
physical reason for phase separation and interfacial tension in
multiphase systems.

So far, several LB multiphase models have been proposed. Among them,
the four well-known models are the chromodynamic model by Gunstensen
et al. \cite{chromodynamic model}, the pseudo-potential model by
Shan and Chen (SC) \cite{shanchen}, the free-energy model by Swift
et al. \cite{Yeomans}, and the HCZ model by He, Chen, and Zhang
\cite{HCZ model}. The chromodynamic model is developed from the
two-component Lattice Gas Automata (LGA) model originally proposed
by Rothman and Keller \cite{RK model}. In this model, the red and
blue colored particles are employed to represent two different
fluids. Phase separation is achieved through controlling the IPI
based on the color gradient. Similar to the treatment in molecular
dynamics (MD), in SC model, non-local interactions between particles
at neighboring lattice sites are incorporated. The interactions
determine the form of the equation of state (EOS). Phase separation
or mixing is governed by the mechanical instability when the sign of
the IPI is properly chosen. In the free-energy model, besides the
mass and momentum conservation constraints, additional ones are
imposed on the equilibrium distribution function, which makes the
pressure tensor consistent with that of the free-energy functional
of inhomogeneous fluids. In the HCZ model, two distribution
functions are used. The first one is used to compute the pressure
and the velocity fields. The other one is used to track interfaces
between different phases. Molecular interactions, such as the
molecular exclusion volume effect and the intermolecular attraction,
are incorporated to simulate phase separation and interfacial
dynamics.

The aforementioned models have been successfully applied to a wide
variety of multiphase and/or multicomponent flow problems, including
drop breakup \cite{3D-DROPLET BREAK,break-pre-2002}, drop collisions \cite%
{POF-PRE-2005-Abraham}, wetting \cite{SC model--wetting-PRE-2006,SC
model-nonwetting-PRE-2007}, contact line motion \cite%
{contact-line-motion-2004, contact angle shanchen}, chemically reactive
fluids \cite{reactive-Yeomans-2006}, phase separation and phase ordering
phenomena \cite{Yeomans,XGL1,Sofonea-multiphase,Sofonea-1999EurPhysJB}, hydrodynamic instability \cite%
{RT-PRE-1998,ZhangRY}, etc. However, despite this, the current LB
versions for multiphase flows are still subjected to, at least, one
of the following constrains (i) the isothermal constraint (i.e., the
deficiency of temperature dynamic), (ii) the limited density ratio
and temperature range, (iii) the spurious velocities. This paper
addresses mainly the last restriction and the total energy
conservation in practical simulations.

Spurious velocities extensively exist in simulations of the
liquid-vapor system and reach their maxima at the interfacial
regions, indicating deviation from the real physics of a fluid
system. Reducing and eliminating the unphysical velocities are of
great importance to the simulations of multiphase flows. Firstly,
large spurious velocities will lead to numerical instability.
Secondly, the local velocities are small during phase separation and
coarsening. If the spurious currents are too large, then we may not
be able to separate the spurious currents from the real local flows,
which is especially true in the case of phase separation with high
viscosity. Thirdly, for a thermal multiphase system, accurate flow
velocities are required in order to obtain an accurate temperature
field \cite{Yuan-POF-2006}.

In dealing with this issue, extensive efforts have been made during
the past years. Wagner \cite{Wanger-IJMPC-2003} pointed out that the
origin of the spurious currents is due to the incompatibility
between the discretizations of driving forces for the order
parameter and momentum equation. Therefore, he suggested to cure the
spurious velocities by removing the nonideal terms from the pressure
tensor and introducing them as a body force. Sofonea and Cristea et
al. \cite{Sofonea-multiphase,Sofonea-us} presented a finite
difference LB (FDLB) approach and proposed two ways to eliminate the
unwanted currents. In the first way, a high-accuracy numerical
scheme, the flux limiter method is employed to calculate the
convection term of the LB equation. In the second way, a correction
force term is introduced to the LB equation that cancels the
spurious velocities and allows to recover the mass equation
correctly. Shan \cite{Shan-PRE-2006-2008} and Succi et al.
\cite{Succi-PRE-2007} showed that the origin of the spurious
currents are due to the insufficient isotropy in the calculation of
density gradient. Therefore, using the information of the density
field on an extended neighboring $\mathbf{x}+\mathbf{e}_{i}$ of a
given site to construct high order isotropic difference operators,
is the key for the correct discretization of spatial derivatives and
taming the spurious currents in the interface. Yuan et al.
\cite{Yuan-POF-2006} demonstrated that smaller parasitical
velocities and higher density ratio can be achieved using more
realistic EOS in a single-component multiphase LB model. Lee and
Fischer \cite{Lee-2006} reported that the use of the potential form
of the surface tension and the isotropic FD scheme can eliminate
parasitic currents to round-off. Seta and Okui \cite{Seta-us}
composed a more accurate fourth order scheme to calculate the
derivatives in the pressure tensor. This convenient approach reduces
the amplitude of spurious velocities to about
one half of that from the second order scheme. Pooley and Furtado \cite%
{Sten-2008} analyzed the causes of spurious velocities in a
free-energy LB model and provided two improvements. First, by making
a suitable choice of the equilibrium distribution and using the
nine-point stencils (NPS) scheme to calculate derivatives, the
magnitude of spurious velocities can be decreased by an order.
Moreover, a momentum conserving force is presented to further reduce
the spurious velocities. Yeomans et al. \cite{Yeomans-MRT}
identified two sources of the spurious velocities, the long range
effects and the bounce-back boundary conditions, when a single
relaxation time (SRT) LB algorithm is used to solve the hydrodynamic
equations of a binary fluids. Aiming to reduce the unwanted
velocities, they proposed a revised LB method based on a
multiple-relaxation-time (MRT) algorithm.

In this work, we present a thermal LB model for simulating thermal
liquid-vapor system with neglectable spurious velocities. This model
is a further development of the one originally proposed by Watari
and Tsutahara (WT) \cite{Watari} and then developed by Gonnella,
Lamura and Sofonea (GLS) \cite{GLS}. The original WT model works
only for ideal gas. GLS introduced an appropriate IPI force term to
describe van der Waals (VDW) fluids. Here we introduce a windowed
FFT (WFFT) scheme to calculate the convection term and the force
term. The improved model is convenient to compromise the high
accuracy and stability. With the new model, non-conservation problem
of total energy due to spatiotemporal discretizations is much better
controlled and spurious currents in equilibrium interfaces are
significantly damped.

The rest of the paper is structured as follows. In the next section
the thermal LB models for ideal gas and for VDW fluids are briefly
reviewed. In section III we illustrate the necessity of the further
development and detail the usage of the WFFT scheme and its inverse.
Comparisons and analysis of numerical results from different schemes
are presented in section IV, where we will show how the spurious
velocities around linear and curved interfaces can be reduced by the
new model. Finally, in section V, we summarize the results and
suggest directions for future research.

\section{The model}

\subsection{Original WT model for ideal gas system}

The thermal multiphase model is developed from the thermal LB model,
originally, proposed by WT, which is based on a multispeed approach.
In this approach, additional speeds are required and higher order
velocity terms are included in the equilibrium distribution function
to obtain the macroscopic temperature field.

WT model uses the following discrete-velocity-model (DVM) which involves a
set of 33 nondimensionalized velocities
\begin{equation}
\mathbf{v}_{0}=0\text{, }\mathbf{v}_{ki}=v_{k}[\cos (\frac{i-1}{4}\pi )\text{%
,}\sin (\frac{i-1}{4}\pi )]\text{,}  \label{dvm_eq_1}
\end{equation}%
where subscript $k=1$,...,$4$ indicates the $k$-th group of the particle
velocities whose speed is $v_{k}$ and $i=1$,...,$8$ indicates the direction
of particle's speed. In our simulations we set $\mathrm{\ }v_{1}=1.00$,$%
\mathrm{\ }v_{2}=1.90$,$\mathrm{\ }v_{3}=2.90$, and $\mathrm{\ }v_{4}=4.30$.

The distribution function $f_{ki}$, discrete in space and time, evolves
according to a SRT Boltzmann equation
\begin{equation}
\frac{\partial f_{ki}}{\partial t}+\mathbf{v}_{ki}\cdot \frac{\partial f_{ki}%
}{\partial \mathbf{r}}=-\frac{1}{\tau }\left[ f_{ki}-f_{ki}^{eq}\right]
\text{,}  \label{bgk_eq}
\end{equation}%
where $f_{ki}^{eq}$, $\mathbf{r}$, and $\tau $ denote the local
equilibrium distribution function, the spatial coordinate, and the
relaxation time, respectively. $f_{ki}^{eq}$ is expressed as a
series expansion in the local velocity
\begin{eqnarray}
f_{ki}^{eq} &=&\frac{\rho }{{2\pi T}}\exp (-\frac{{v_{ki}^{2}}}{{2T}})\exp (-%
\frac{{u^{2}-2uv_{ki}}}{{2T}})  \notag \\
&=&\rho F_{k}[(1-\frac{{u^{2}}}{{2T}}+\frac{{u^{4}}}{{8T^{2}}})+\frac{{%
v_{ki\varepsilon }u_{\varepsilon }}}{T}(1-\frac{{u^{2}}}{{2T}})  \notag \\
&&+\frac{{v_{ki\varepsilon }v_{ki\pi }u_{\varepsilon }u_{\pi }}}{{2T^{2}}}(1-%
\frac{{u^{2}}}{{2T}})+\frac{{v_{ki\varepsilon }v_{ki\pi }v_{ki\eta
}u_{\varepsilon }u_{\pi }u_{\eta }}}{{6T^{3}}}  \notag \\
&&+\frac{{v_{ki\varepsilon }v_{ki\pi }v_{ki\eta }v_{ki\lambda
}u_{\varepsilon }u_{\pi }u_{\eta }u_{\lambda }}}{{24T^{4}}}]+\cdot \cdot
\cdot \text{,}  \label{feq}
\end{eqnarray}%
where the weight factors are
\begin{eqnarray}
F_{k} &=&\frac{{1}}{{v_{k}^{{2}}(v_{k}^{{2}}-v_{k+1}^{{2}})(v_{k}^{{2}%
}-v_{k+2}^{{2}})(v_{k}^{{2}}-v_{k{+3}}^{{2}})}}{[48}T^{{4}}-{6(}v_{k+1}^{2}{+%
}v_{k+2}^{{2}}{+}v_{k+3}^{{2}}{)}T^{{3}}  \notag \\
&&+{(}v_{k+1}^{{2}}v_{k+2}^{{2}}{+}v_{k+2}^{{2}}v_{k+3}^{{2}}{+}v_{k+3}^{{2}%
}v_{k+1}^{{2}}{)}T^{{2}}-\frac{{v_{k+1}^{{2}}v_{k+2}^{{2}}v_{k+3}^{{2}}}}{{4}%
}T{]}\text{{,}}  \label{FK}
\end{eqnarray}%
\begin{equation}
F_{0}={1}-{8(}F_{1}{+}F_{2}{+}F_{3}{+}F_{4}{)}\text{,}  \label{F0}
\end{equation}%
with
\begin{equation}
\left\{ {k+l}\right\} =\left\{
\begin{array}{cc}
k+l\text{,} & k+l\leq 4\text{;} \\
k+l-4\text{,} & k+l>4\text{;}%
\end{array}%
l\in \{1,2,3\}\text{.}\right.   \label{K+L}
\end{equation}

Hydrodynamic quantities, such as density, velocity, and temperature
are determined from the following moments
\begin{equation}
\rho =\sum_{ki}f_{ki}^{eq}\text{,}  \label{n_eq}
\end{equation}%
\begin{equation}
\rho \mathbf{u}=\sum_{ki}\mathbf{v}_{ki}f_{ki}^{eq}\text{,}  \label{nu_eq}
\end{equation}%
\begin{equation}
\rho T=\sum_{ki}\frac{1}{2}(\mathbf{v}_{ki}-\mathbf{u})^{2}f_{ki}^{eq}\text{.%
}  \label{p_eq}
\end{equation}

The combination of the above DVM and the general FD scheme with
first-order forward in time and second-order upwinding in space
composes the original FDLB model by WT. In the FDLB model, particle
velocities are independent from the lattice configuration. As a
result, higher-order numerical schemes can be used to reduce the
numerical viscosity and to enhance the stability of the model. This
is of great importance to LB simulations, especially in phase
separation studies, where long lasting simulations are needed to
establish the growth properties.

\subsection{GLS model for multiphase system}

WT model can be applied to compressible flows with small Mach number
and the revised version \cite{Xu-compressible-2} extends it to
compressible flows with high Mach number due to better numerical
stability. Nevertheless, neither the original one nor the improved
one has the ability to describe multiphase flows, since both models
lead to the ideal EOS only, which do not support thermodynamical
two-phase state.

Fortunately, by incorporating a forcing term, the improved model can
be applied to thermal liquid-vapor systems. Compared to isothermal
models, the variable temperature that the GLS model can be
implemented is of great importance, since thermal effects are
ubiquitous and sometimes dominant in an important class of flows
\cite{Fun-Multiphase}. Examples are referred to boiling
\cite{Boiling}, distillation, as well as the dynamics of phase
separation \cite{Onuki-book, Onuki-PRL-PRE-EPL-PRE}, where the
freedom in temperature limits the rate of phase separation and
induces different rheological and morphological behaviors. Dynamic
effects of temperature can not be considered in isothermal models,
therefore, most studies have been restricted to either isothermal
systems or the systems where effects of temperature dynamics are
negligible.

The forcing term introduced by GLS is added into the right-hand side
(RHS) of Eq. (2)
\begin{equation}
\frac{\partial f_{ki}}{\partial t}+\mathbf{v}_{ki}\cdot \frac{\partial f_{ki}%
}{\partial \mathbf{r}}=-\frac{1}{\tau }\left[ f_{ki}-f_{ki}^{eq}\right]
+I_{ki}\text{,}  \label{IKI-BGK}
\end{equation}%
where $I_{ki}$ takes the following form,
\begin{equation}
I_{ki}=-[A+B_{\alpha }(v_{ki\alpha }-u_{\alpha })+(C+C_{q})(v_{ki\alpha
}-u_{\alpha })^{2}]f_{ki}^{eq}\text{.}  \label{iki}
\end{equation}%
$I_{ki}$ in Eq. (10) is introduced to control the equilibrium properties of
the liquid-vapor systems and allows to recover the following equations for
VDW fluids \cite{GLS},
\begin{equation}
\partial _{t}\rho +\partial _{\alpha }(\rho u_{\alpha })=0\text{,}
\label{mass}
\end{equation}%
\begin{equation}
\partial _{t}(\rho u_{\alpha })+\partial _{\beta }(\rho u_{\alpha }u_{\beta
}+\Pi _{\alpha \beta }-\sigma _{\alpha \beta })=0\text{,}  \label{moment}
\end{equation}%
\begin{equation}
\partial _{t}e_{T}+\partial _{\alpha }[e_{T}u_{\alpha }+(\Pi _{\alpha \beta
}-\sigma _{\alpha \beta })u_{\beta }-\kappa _{T}\partial _{\alpha }T]=0\text{%
,}  \label{energy}
\end{equation}%
where
\begin{equation}
\Pi _{\alpha \beta }=P^{w}\delta _{\alpha \beta }+\Lambda _{\alpha \beta }%
\text{,}  \label{NVS}
\end{equation}%
\begin{equation}
\sigma _{\alpha \beta }=\eta (\partial _{\alpha }u_{\beta }+\partial _{\beta
}u_{\alpha }-\partial _{\gamma }u_{\gamma }\delta _{\alpha \beta })+\zeta
\partial _{\gamma }u_{\gamma }\delta _{\alpha \beta }\text{,}  \label{DT}
\end{equation}%
and
\begin{equation}
e_{T}=\rho T-9\rho ^{2}/8+K\left\vert \nabla \rho \right\vert ^{2}/2+\rho
u^{2}/2\text{,}
\end{equation}%
represent the non-viscous stress, the dissipative tensor, and the
total energy density, respectively. $\kappa _{T}$, $\eta $, and
$\zeta $ are heat conductivity, shear, and bulk viscosities. $P^{w}$
and $\Lambda _{\alpha \beta }$ in Eq. (\ref{NVS}) are the VDW EOS
and the contribution of density gradient to pressure tensor, which
have the following expressions\qquad
\begin{equation}
P^{w}=3\rho T/(3-\rho )-9\rho ^{2}/8\text{,}  \label{EOS}
\end{equation}%
\begin{equation}
\Lambda _{\alpha \beta }=M\partial _{\alpha }\rho \partial _{\beta }\rho
-[\rho T\partial _{\gamma }\rho \partial _{\gamma }(M/T)]\delta _{\alpha
\beta }-M(\rho \nabla ^{2}\rho +\left\vert \nabla \rho \right\vert
^{2}/2)\delta _{\alpha \beta }\text{.}
\end{equation}%
The expression $M=K+HT$ allows a dependence of the surface tension on
temperature, where $K$ is the surface tension coefficient and $H$ is a
constant.

In order to recover Eqs. (\ref{mass})-(\ref{energy}), five
constraints are imposed on the forcing term, which make coefficients
in Eq. (\ref{iki}) as the following form
\begin{equation}
A=-2(C+C_{q})T\text{,}  \label{AAA}
\end{equation}%
\begin{equation}
B_{\alpha }=\frac{1}{\rho T}[\partial _{\alpha }(P^{w}-\rho T)+\partial
_{\beta }\Lambda _{\alpha \beta }-\partial _{\alpha }(\zeta \partial
_{\gamma }u_{\gamma })]\text{,}  \label{BBB}
\end{equation}%
\begin{eqnarray}
C &=&\frac{1}{2\rho T^{2}}\{(P^{w}-\rho T)\partial _{\gamma }u_{\gamma
}+\Lambda _{\alpha \beta }\partial _{\alpha }u_{\beta }-(\zeta \partial
_{\gamma }u_{\gamma })\partial _{\alpha }u_{\alpha }  \notag \\
&&+\frac{9}{8}\rho ^{2}\partial _{\gamma }u_{\gamma }+K[-\frac{1}{2}%
(\partial _{\gamma }\rho )(\partial _{\gamma }\rho )(\partial _{\alpha
}u_{\alpha })  \notag \\
&&-\rho (\partial _{\gamma }\rho )(\partial _{\gamma }\partial _{\alpha
}u_{\alpha })-(\partial _{\gamma }\rho )(\partial _{\gamma }u_{\alpha
})(\partial _{\alpha }\rho )]\}\text{,}
\end{eqnarray}%
\begin{equation}
C_{q}=\frac{1}{2\rho T^{2}}\partial _{\alpha }[2q\rho T(\partial _{\alpha
}T)]\text{.}  \label{CCQQ}
\end{equation}%
It is worth noting that in this model the Prandtl number $\Pr =\eta /\kappa
_{T}=\tau /2(\tau -q)$ can be changed by adjusting the parameter $q$ in the
term $C_{q}$.

\section{Thermal LB model based on the WFFT approach}

In this section, we present our contribution to the thermal multiphase LB
model: spatial derivatives in the convection term $\mathbf{v}_{ki}\cdot
\partial f_{ki}/\partial \mathbf{r}$ and in the forcing term $I_{ki}$, are
calculated via the WFFT algorithm and its inverse.

To illustrate the necessity, we present simulation results for a thermal
phase separation process by various numerical schemes. Here the time
derivative is calculated using the first-order forward Euler FD scheme. The
spatial derivatives in $I_{ki}$ are calculated using the second-order
central difference scheme. Spatial derivatives in the convection term $%
\mathbf{v}_{ki}\cdot \partial f_{ki}/\partial \mathbf{r} $ are calculated
using various schemes listed as follows:

\subsection{second-order central difference scheme}

Let $J-1$, $J$ and $J+1$ be three successive nodes of the one dimensional
lattice. Using the second-order central difference scheme to discretize the
convention term, Eq. (\ref{IKI-BGK}) can be rewritten in a conservative form
\begin{equation}
f_{ki,J}^{n+1}=f_{ki,J}^{n}-\frac{{c_{ki\alpha }}}{2}%
(f_{ki,J+1}^{n}-f_{ki,J-1}^{n})-\frac{{\Delta t}}{\tau }%
(f_{ki,J}^{n}-f_{ki,J}^{eq,n})+I_{ki,J}^{n}\Delta t\text{,}  \label{2nd-CD}
\end{equation}%
where ${\Delta t}$ and ${c_{ki\alpha }=v}_{ki\alpha }\Delta t/\Delta
r_{\alpha }{\ }$are the time step and the Courant-Friedrichs-Levy (CFL)
number.

\subsection{Lax-Wendroff (LW) scheme}

Compared with the second-order central difference scheme, the LW
scheme contributes a dissipation term, which is in favor of the
numerical stability. Then, by using this scheme, Eq. (\ref{IKI-BGK})
can be formulated as
\begin{eqnarray}
f_{ki,J}^{n+1} &=&f_{ki,J}^{n}-\frac{{c_{ki\alpha }}}{2}%
(f_{ki,J+1}^{n}-f_{ki,J-1}^{n})+\frac{{c_{ki\alpha }^{2}}}{2}%
(f_{ki,J+1}^{n}-2f_{ki,J}^{n}+f_{ki,J-1}^{n})  \notag \\
&&-\frac{{\Delta t}}{\tau }(f_{ki,J}^{n}-f_{ki,J}^{eq,n})+I_{ki,J}^{n}\Delta
t\text{.}  \label{ML-}
\end{eqnarray}

\subsection{Modified-LW (MLW) scheme}

As we know, the LW scheme is very dissipative and has a strong ``
smoothing effect". Obviously, it is not favorable to recover the
sharp interface in the multiphase system. To further improve the
numerical accuracy, the modified partial differential equation
(MPDE) remainder after discretizing with Eq. (\ref{ML-}) is derived
\cite{Liu-MPDE}
\begin{equation}
R=-\frac{{v_{ki\alpha }(1-c_{ki\alpha }^{2})}}{6}\Delta r_{\alpha }^{2}\frac{%
{\partial ^{3}f}}{{\partial r^{3}}}-\frac{{v_{ki\alpha }c_{ki\alpha
}(1-c_{ki\alpha }^{2})}}{8}\Delta r_{\alpha }^{3}\frac{{\partial ^{4}f}}{{%
\partial r^{4}}}+\cdots \text{.}  \label{MPDE}
\end{equation}%
It is clear that the first and the second terms in the RHS of Eq. (\ref{MPDE}%
) correspond to the third-order dispersion error $R_{3}$ and the
fourth-order dissipation error $R_{4}$, respectively. Therefore, we can add
the dispersion term into the RHS of Eq. (\ref{IKI-BGK}) to compensate the
dispersion error
\begin{eqnarray}
\frac{{\partial f_{ki}}}{{\partial t}}+\mathbf{v}_{ki}.\frac{\partial }{{%
\partial \mathbf{r}}}f_{ki} &=&-\frac{{1}}{\tau }(f_{ki}-f_{ki}^{eq})+I_{ki}
\notag \\
&&{+}\frac{{v_{ki\alpha }(1-c_{ki\alpha }^{2})}}{6}\Delta r_{\alpha }^{2}%
\frac{{\partial ^{3}f}}{{\partial r^{3}}}\text{.}  \label{dispersion}
\end{eqnarray}%
Furthermore, we can add the dissipation term into RHS of Eq. (\ref{dispersion}%
). Using the 2nd-CD scheme to discrete $R_{3}$ and $R_{4}$ gives
\begin{equation}
\bar{R}_{3}=\frac{{c_{ki\alpha }(1-c_{ki\alpha }^{2})}}{{12}}%
(f_{ki,J+2}^{n}-2f_{ki,J+1}^{n}+2f_{ki,J-1}^{n}-f_{ki,J-2}^{n})\text{,}
\label{R3}
\end{equation}%
and
\begin{equation}
\bar{R}_{4}=\frac{{c_{ki\alpha }^{2}(1-c_{ki\alpha }^{2})}}{8}%
(f_{ki,J+2}^{n}-4f_{ki,J+1}^{n}+6f_{ki,J}^{n}-4f_{ki,J-1}^{n}+f_{ki,J-2}^{n})%
\text{.}  \label{R4=}
\end{equation}%
The bars above $R_{3}$ and $R_{4}$ indicate that they are discretized. If only $\bar{R}%
_{3}$ is added into the RHS of Eq. (\ref{ML-}), for convenience of
description, we refer to this scheme as MLW1. If both $\bar{R}_{3}$ and $%
\bar{R}_{4}$ are added into the RHS of Eq. (\ref{ML-}), then a more accurate
LB equation is obtained, and we refer to this scheme as MLW2.

\subsection{Flux limiter (FL) scheme}

\begin{figure}[tbp]
\center{\epsfig{file=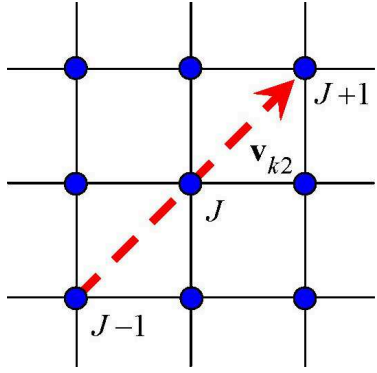,bbllx=0pt,bblly=0pt,bburx=197pt,bbury=197pt,
width=0.3\textwidth,clip=}}
\caption{(Color online) Characteristic line on the square lattice for the
direction $i=2$. }
\end{figure}
The FL scheme has been widely employed by Sofonea et al. \cite%
{Sofonea-multiphase,Sofonea-us} to reduce the spurious velocities and to
improve the numerical stability in liquid-vapor systems. Figure 1 shows the
characteristic line on the square LB lattice for direction $i=2.$ When using
this approach to compute the convective term along the characteristic line,
Eq. (\ref{IKI-BGK}) becomes
\begin{eqnarray}
f_{ki,J}^{n+1} &=&f_{ki,J}^{n}-\frac{{v_{k}\Delta t}}{{A_{i}\Delta r_{\alpha
}}}[F_{ki,J+1/2}^{n}-F_{ki,J-1/2}^{n}]  \label{FL} \\
&&-\frac{{1}}{\tau }(f_{ki,J}^{n}-f_{ki,J}^{n,{eq}})\Delta
t+I_{ki,J}^{n}\Delta t\text{,}
\end{eqnarray}%
with
\begin{equation}
A_{i}=\left\{
\begin{array}{cc}
1\text{,} & i\in \{1,3,5,7\}\text{;} \\
\sqrt{2}\text{,} & i\in \{2,4,6,8\}\text{.}%
\end{array}%
\right.
\end{equation}%
$F_{ki,J+1/2}^{n}$ and $F_{ki,J-1/2}^{n}$ in Eq. (\ref{FL}) are two fluxes,
which are defined as
\begin{equation}
F_{ki,J+1/2}^{n}=f_{ki,J}^{n}+\frac{1}{2}(1-\frac{{v_{k}\Delta t}}{{%
A_{i}\Delta r_{\alpha }}})[f_{ki,J+1}^{n}-f_{ki,J}^{n}]\psi (\theta
_{ki,J}^{n})\text{,}
\end{equation}%
\begin{equation}
F_{ki,J-1/2}^{n}=F_{ki,(J-1)+1/2}^{n}\text{,}
\end{equation}%
where the flux limiter $\psi (\theta _{ki,J}^{n})$ is expressed as a
smooth function
\begin{equation}
\theta _{ki,J}^{n}=\frac{f_{ki,J}^{n}-f_{ki,J-1}^{n}}{%
f_{ki,J+1}^{n}-f_{ki,J}^{n}}\text{.}
\end{equation}%
In particular, if $\psi (\theta _{ki,J}^{n})=0$, it corresponds to the
first-order upwind scheme and $\psi (\theta _{ki,J}^{n})=1$ to the LW
scheme. A wide choice of flux limiters can work well with LB models. In this
work, we will use the monitorized central difference (MCD) FL, which is most
widely used by Sofonea et al.
\begin{equation}
\psi (\theta _{ki,J}^{n})=\left\{
\begin{array}{cc}
0\text{,} & \theta _{ki,J}^{n}\leq 0\text{,} \\
2\theta _{ki,J}^{n}\text{,} & 0<\theta _{ki,J}^{n}\leq 1/3\text{,} \\
(1+\theta _{ki,J}^{n})/2\text{,} & 1/3<\theta _{ki,J}^{n}\leq 3\text{,} \\
2\text{,} & 3<\theta _{ki,J}^{n}\text{.}%
\end{array}%
\right.
\end{equation}

\subsection{NPS scheme}

Recently, a new scheme, named the NPS scheme, has been widely used
to calculate the spatial derivatives by many scholars so as to
ensure higher isotropy and to reduce spurious velocities
\cite{Yuan-POF-2006,Sten-2008,
Yeomans-MRT,JCP-Lee-2005,Gonnella-2009}. The general choice of
stencils for calculating the derivatives and the laplacian are
\begin{eqnarray}
\bar{\partial}_{x} &=&\frac{1}{{\Delta x}}\left[
\begin{array}{ccc}
-B & 0 & B \\
-A & 0 & A \\
-B & 0 & B%
\end{array}%
\right]   \notag \\
&=&\partial _{x}+\frac{1}{6}\Delta x^{2}\partial _{x}^{3}+2B\Delta
x^{2}\partial _{x}\partial _{y}^{2}+\cdots \text{,}
\end{eqnarray}%
and
\begin{eqnarray}
\bar{\nabla}^{2} &=&\frac{1}{{\Delta x^{2}}}\left[
\begin{array}{ccc}
F & E & F \\
E & -4(E+F) & E \\
F & E & F%
\end{array}%
\right]   \notag \\
&=&\nabla ^{2}+\frac{{\Delta x^{2}}}{{12}}(\partial _{x}^{4}+\partial
_{y}^{4})+F\Delta x^{2}\partial _{x}^{2}\partial _{y}^{2}+\cdots \text{,}
\end{eqnarray}%
with $2A+4B=1$ and $E+2F=1$ to keep consistency between the continuous and
discrete operators. The bars above $\partial _{x}$ and $\nabla ^{2}$
represent that they are discrete operators. The central entry denotes the
lattice node at which the derivative is calculated, and the remaining
entries are the eight neighbor nodes around the central one. $B$ and $F$ are
two free parameters that are chosen to minimize the spurious velocities. A
large amount of numerical tests indicate that the best choice is $B=1/12$ and $%
F=1/6$ in the GLS model. In our simulations, both the convection
term and the forcing term are calculated by this way.

\begin{figure}[tbp]
\center{\epsfig{file=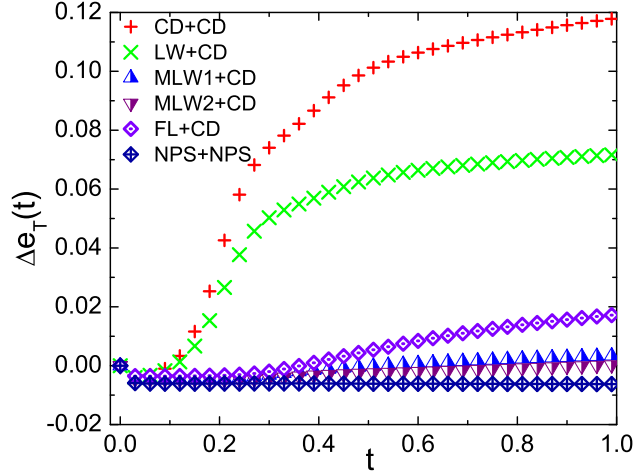,bbllx=7pt,bblly=4pt,bburx=566pt,bbury=430pt,
width=0.5\textwidth,clip=}} \caption{(Color online) Variations of
total energy $\Delta e_{T}(t)=e_{T}(t)-e_{T}(0)$ for a phase
separating process with various numerical schemes. }
\end{figure}

Next, we conduct simulations of a thermal phase separation process
with numerical schemes listed above. Initial conditions of our test
are chosen as
\begin{equation}
(\rho \text{, }T\text{, }u\text{, } v)=(1+\Delta \text{,}0.85\text{,
}0.0\text{, }0.0)\text{,}
\end{equation}%
where $\Delta $ is a random density with an amplitude $0.001$ and
can be regarded as incipient nuclei in the density field. Periodical
boundary conditions (PBC) are imposed on a square lattice with
$N_{x}=N_{y}=128$.
Unless otherwise stated, the remaining parameters are $\Delta x=\Delta y=1/256$, $%
\Delta t=10^{-5}$, $\tau =10^{-2}$, $K=10^{-5}$, $H=0$, $\zeta =0$,
$q=-0.004$,
throughout our simulations. Figure 2 shows the variations of total energy $%
\Delta e_{T}(t)=e_{T}(t)-e_{T}(0)$ for the phase separating process
with various numerical schemes. The legend in each case is composed
of two parts, `A'+`B', where `A' is `CD', `LW', `MLW1', `MLW2', `FL'
and `NPS' and it shows the scheme to calculate the convection term;
`B' is `CD' and `NPS' and it shows the scheme to calculate the
forcing term. Figure 2 demonstrates that the total energy density
$e_{T}(t)$ is not conservative in simulations even though it is in
theoretical analysis. Further survey of these results indicates that
the derivation $\Delta e_{T}(t)$ decreases by increasing the
accuracy of scheme. Therefore, we conclude that the non-conservation
of total energy is mainly due to the spatial discretization errors.

\subsection{The WFFT approach}

To overcome the problem of energy non-conservation, a new algorithm
based on WFFT is proposed. This approach is especially powerful for
periodic system and also provides spatial spectral information on
hydrodynamic quantities. Moreover, with this approach, higher-order
derivatives and fractional-order derivatives can be computed in a
convenient way.

For the sake of clarity, we start with the definition of Fourier transform
of $f(x_{j})$
\begin{equation}
\widetilde{f}(k)=\Delta x\sum_{j=0}^{N-1}f(x_{j})e^{-\mathbf{i}kx_{j}}\text{,%
}  \label{FT}
\end{equation}%
and its inverse
\begin{equation}
f(x_{j})=\frac{1}{L}\sum_{n=-N/2}^{N/2-1}\widetilde{f}(k)e^{\mathbf{i}kx_{j}}%
\text{,}  \label{IFT}
\end{equation}%
where $k$ is the module of wave vector $\mathbf{k}$, $\mathbf{i}$ is the
imaginary unit, and $\widetilde{f}(k)$ stands for the Fourier transform of a
spatial function $f(x)$. In Eq. (\ref{IFT}), $k\mathbf{=}2\pi n/L$, and $%
L=N\Delta x$ is the length of the system divided into $N$ equal segments.
The above two equations are exactly correct when $N$ is infinitely large or $%
\Delta x$ is infinitely small. A general theorem of derivative based on FFT
states that \cite{sxwlff,Spectral-Methods-Book,Plasma-Book}
\begin{equation}
\widetilde{f^{\prime }}(k)=\mathbf{i}k\times \widetilde{f}(k)\text{,}
\label{FFT}
\end{equation}%
where $\widetilde{f^{\prime }}(k)$ is the Fourier transform of $f^{\prime
}(x)$. The theorem suggests a way to calculate spatial derivative $f^{\prime
}(x)$, as shown in Fig. 3. Firstly, transform $f(x)$ in real space into $%
\widetilde{f}(k)$ in reciprocal space; then, multiply
$\widetilde{f}(k)$ with $\mathbf{i}k$; finally, take the inverse
Fourier transform (IFT) of $\widetilde{f^{\prime }}(k)$, the spatial
derivative $f^{\prime }(x)$ can be obtained.

\begin{figure}[tbp]
\center{\epsfig{file=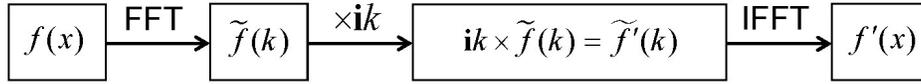,bbllx=0pt,bblly=0pt,bburx=520pt,bbury=49pt,
width=0.75\textwidth,clip=}}
\caption{A possible flow chart for differential operator using the FFT
scheme and its inverse (IFFT).}
\end{figure}

The approach mentioned above has excellent accuracy properties, typically
well beyond that of standard discretization schemes. In principle, it gives
the exact derivative with infinite order accuracy if the function is
infinitely differentiable \cite{Spectral-Methods-Book,Plasma-Book,
Spectral-Methods-Book-2,PRL-FFT-1971}, which is another advantage of FFT
scheme compared to the FD scheme. In our manuscript, using this virtue, the
FFT scheme is designed to approximate the true spatial derivatives, as a
result, to eliminate spurious velocities and to guarantee energy
conservation.

However, the trouble in proceeding in this manner is that, in many
cases, it is difficult to ensure that infinite differentiability
condition is satisfied. For example, the Sod shock tube problem
\cite{Sod} contains the shock wave, the rarefaction wave and the
contact discontinuity.  Then the derivative of hydrodynamic
quantity, $\rho^{\prime}(x)$ or $T^{\prime}(x)$ has a discontinuity
as the same character as the square wave (see Fig.6 for more
details). Then the discontinuity will induce oscillations, known as
the Gibbs phenomenon. The Gibbs phenomenon influences the accuracy
of the FFT not only in the neighborhood of the point of singularity,
but also over the entire computational domain. More importantly,
sometimes, it will cause numerical instability. For example, for the
problems shown in Figs. 9 and 17 (in Sec. IV), the above approach is
unstable due to the Gibbs phenomenon. Recently, there is a trend to
use smoothing procedures which attenuate higher-order Fourier
coefficients to avoid or at least to reduce these oscillations
(i.e., WFFT method) \cite{Spectral-Methods-Book,
Spectral-Methods-Book-2,Fliters,JCP-2006}. A straightforward and
convenient way to attenuate the higher-order Fourier coefficients is
to multiply each Fourier coefficients by a smoothing factor (filter)
$\sigma_{k}$, such as the Lanzos filter, raised cosine filter,
sharpened raised cosine filter and exponential cutoff filter, as
listed in Refs. \cite{Spectral-Methods-Book,Fliters,JCP-2006}.

In the present study, based on Taylor series expansion of wave
number $k$, we present a way to construct smoothing factors.
Firstly, we expand $k$ in Taylor series
\begin{eqnarray}
k &=&\frac{\arcsin [\sin (k\Delta x/2)]}{\Delta x/2}  \notag \\
&=&\frac{1}{\Delta x/2}[\sin (k\Delta x/2)+\frac{1}{6}\sin ^{3}(k\Delta x/2)+%
\frac{3}{40}\sin ^{5}(k\Delta x/2)+\frac{5}{112}\sin ^{7}(k\Delta x/2)+...]
\notag \\
&=&\frac{1}{\Delta x/2}\sum_{n=0}^{\infty }\frac{\Gamma (n/2)\delta
_{0,\Theta (n)}\varepsilon (-1+n)}{\sqrt{\pi }n\Gamma (\frac{n+1}{2})}\sin
^{n}(k\Delta x/2)\text{,}  \label{K-TaylorSeries}
\end{eqnarray}%
where $\Gamma (n)=\int_{0}^{\infty }t^{n-1}e^{-t}dt$ is the Gamma function, $%
\Theta (n)=Mod[-1+n,2]$ is the Mod function, and $\varepsilon (-1+n)$ is the
unit step function. Thus, in order to damp the Gibbs oscillations, or in
order to filter out more high frequency waves, $k$ may take the form of an
appropriately truncated Taylor series expansion of sin$(k\Delta x/2)$. For
example, $k$ may take the following forms
\begin{equation}
k_{1}\mathbf{=}\frac{\sin (k\Delta x/2)}{\Delta x/2}\text{,}  \label{K1}
\end{equation}%
\begin{equation}
k_{2}\mathbf{=}k_{1}+\frac{\sin ^{3}(k\Delta x/2)/6}{\Delta x/2}\text{,}
\label{K2}
\end{equation}%
\begin{equation}
k_{3}\mathbf{=}k_{2}+\frac{3\sin ^{5}(k\Delta x/2)/40}{\Delta x/2}\text{,}
\label{K3}
\end{equation}%
and
\begin{equation}
k_{4}\mathbf{=}k_{3}+\frac{5\sin ^{7}(k\Delta x/2)/112}{\Delta
x/2}\text{,} \label{K4}
\end{equation}%
where $k_{1}$ is consistent with the one used in Ref.
\cite{Shu-ICASE Report}. Some simple derivations indicate that the
above approach with $k_{1}$, $k_{2} $, $k_{3}$, and $k_{4}$ has a
second-order, fourth-order, sixth-order, and eighth-order accuracy
in space, respectively (see Appendix for more details).

\begin{figure}[tbp]
\center{\epsfig{file=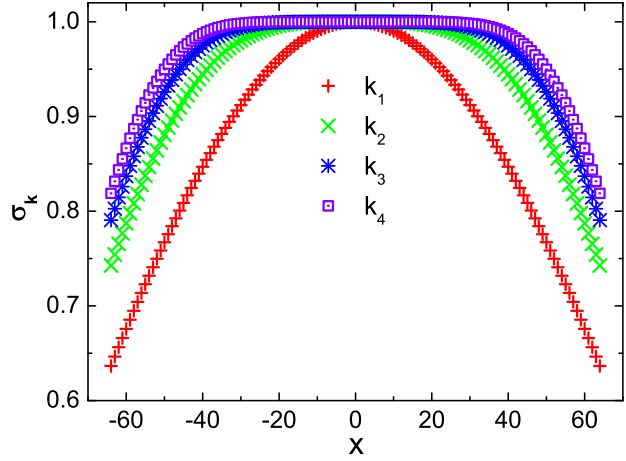,bbllx=11pt,bblly=4pt,bburx=572pt,bbury=420pt,
width=0.5\textwidth,clip=}}
\caption{Smoothing factors $\protect\sigma _{k}$ for $k_{1}$, $k_{2} $, $%
k_{3}$, and $k_{4}$ with $N=128$.}
\end{figure}
\begin{figure}[tbp]
\center{\epsfig{file=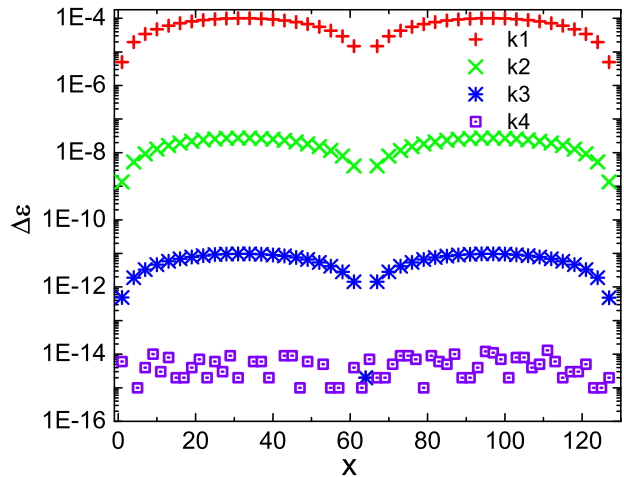,bbllx=6pt,bblly=5pt,bburx=573pt,bbury=445pt,
width=0.5\textwidth,clip=}} \caption{(Color online) Absolute errors
for the computed first-order derivative of the test function
$f(x)=sin(x)$ with the WFFT algorithm. }
\end{figure}
\begin{figure}[tbp]
\center{\epsfig{file=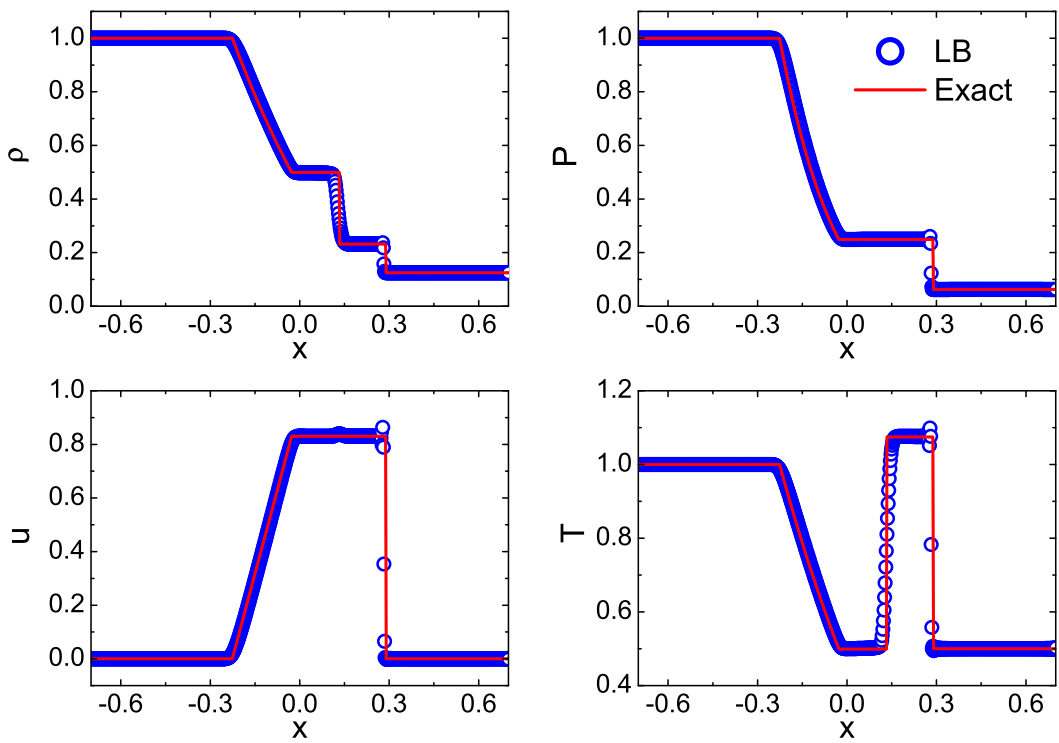,bbllx=10pt,bblly=3pt,bburx=571pt,bbury=402pt,
width=0.85\textwidth,clip=}} \caption{(Color online) Comparisons
between LB results and exact ones for the one-dimensional modified
Sod problem at $t=0.2$. The simulation results (circles) are from
the WT model with the WFFT scheme, where the lower-order filter
$k_{1}$ is used. }
\end{figure}
\begin{figure}[tbp]
\center{\epsfig{file=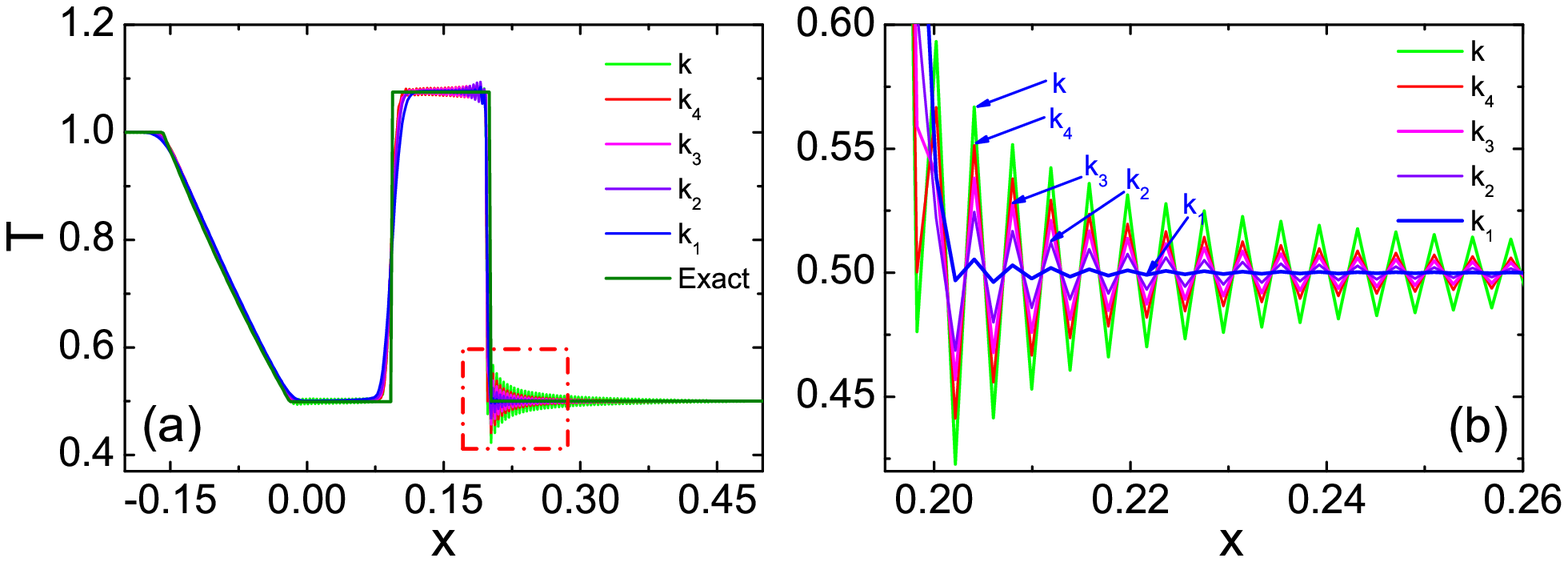,bbllx=2pt,bblly=3pt,bburx=557pt,bbury=207pt,
width=0.85\textwidth,clip=}} \caption{(Color online) Temperature
profiles for the modified Sod shock problem obtained from the WFFT
schemes with $k$, $k_{1}$, $k_{2}$, $k_{3}$, and $k_{4}$ in (a). The
portion in square of (a) is enlarged in (b) for a closing view.}
\end{figure}

Therefore, smoothing factor for $k_{1}$ takes the following form
\begin{equation}
\sigma _{1}=\frac{k_{1}}{k}=\frac{\sin (n\pi /N)}{n\pi /N}\text{, }n=-N/2%
\text{,}...\text{,}N/2\text{.}  \label{Fliter}
\end{equation}
Smoothing factors $\sigma _{k}$ for $k_{2}$, $k_{3}$, and $k_{4}$
can be calculated in a similar way and are represented in Fig. 4. It
is clear that the lower-order smoothing factors $k_{1}$ and $k_{2}$,
filter out more high frequency waves, and may result in excessively
smeared approximations, which are unfaithful representations of the
truth physics. On the other hand, the higher-order smoothing factors
$k_{3}$ and $k_{4}$, reserve more higher frequency waves, but may
not damp the Gibbs phenomenon (see Fig. 7 for more details), then
cause numerical instability. The smoothing factors should survive
the dilemma of stability versus accuracy. In other words, they
should be minimal but make the evolution stable at the same time.

As a simple test, using the WFFT algorithm, the derivative of a
infinite differentiable function $f(x)=sin(x)$, is calculated with
$k_{1}$, $k_{2}$, $k_{3}$, $k_{4}$, and plotted in Fig. 5. It is
clearly seen that when $k_{4}$ is used, the errors reduce to
round-off. As another test, the validity of the WFFT scheme is
verified by the modified Sod shock tube with higher pressure ratio.
For the problem considered, the initial condition is described by
\begin{equation}
\left\{
\begin{array}{l}
\left. {(\rho }\text{{, }}{u}\text{{, }}{v}\text{{,
}}{T)}\right\vert
_{L}=(1.0\text{, }0.0\text{, }0.0\text{, }1.0)\text{{,} \ \ \ }x\leq 0\text{;%
} \\
\left. {(\rho }\text{{, }}{u}\text{{, }}{v}\text{{,
}}{T)}\right\vert
_{R}=(0.125\text{, }0.0\text{, }0.0\text{, }0.5)\text{, \ }x>0\text{.}%
\end{array}%
\right.
\end{equation}%
Subscripts ``$L$" and ``$R$" indicate macroscopic variables at the
left and right sides of the discontinuity. The size of grid is
$\Delta x=\Delta y=2.5\times10^{-3}$, time step is $\Delta
t=10^{-5}$, and relaxation time is $\tau =3\times10^{-4}$. Figure 6
shows the computed density, pressure, velocity, and temperature
profiles at $t=0.2$, where the circles are for simulation results
and solid lines are for analytical solutions. The two sets of
results have a satisfying agreement. Figure 7 shows the temperature
profiles obtained from the WFFT schemes with $k$, $k_{1}$, $k_{2}$,
$k_{3}$, and $k_{4}$ in (a) and local details of the part near the
shock wave in (b). One can see that higher-order filters, such as
$k_{3}$ and $k_{4}$, have higher accuracy in smooth regions, but
cannot refrain the Gibbs phenomenon in unsmooth regions effectively.
The lower-order filters, although are too dissipative, can damp the
spurious oscillations to neglectable scale, which are capable of
shock capturing. Therefore, it should be noted that, for flows
without shock waves and/or discontinuities, the WFFT scheme with
higher-order filter is stable, valid and appropriate. While for the
compressible flows with shock waves and/or discontinuities, the WFFT
scheme with lower-order filter is a more appropriate choice. In the
present study, we focus on the liquid-vapor systems without shock
waves and strong discontinuities. Therefore, the WFFT schemes with
higher-order filters are used.

\begin{figure}[tbp]
\center {
\epsfig{file=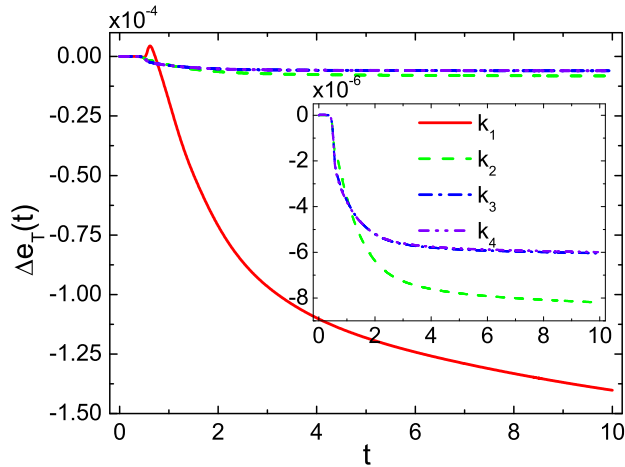,bbllx=7pt,bblly=2pt,bburx=540pt,bbury=405pt,
width=0.5\textwidth,clip=}} \caption{(Color online) Variations of
total energy $\Delta e_{T}(t)$ for the
phase separating process described in Fig. 2, obtained from the WFFT schemes with $%
k_{1}$, $k_{2}$, $k_{3}$, and $k_{4}$. The ones obtained from the
WFFT schemes with $k_{2}$, $k_{3}$, and $k_{4}$ are enlarged in the
inset.}
\end{figure}
For comparisons, we verify the proposed FFT algorithm with the same
problem described in Fig. 2 and display variations of total energy
$\Delta e_{T}(t)$ obtained from WFFT schemes with $k_{1}$, $k_{2}$, $k_{3}$, and $%
k_{4}$ in Fig. 8. It is found that, for each case, $\Delta e_{T}(t)$
oscillates at the beginning, then approaches a nearly constant value. Behaviors of $%
\Delta e_{T}(t)$ can be interpreted as follows. At the beginning of
phase separation, the fluids separate spontaneously into small
regions with higher and lower densities, and more liquid-vapor
interfaces appear. As a result, spacial discretization errors in Eq.
(10) induced by interfaces arrive at their maxima that account for
the initial oscillations. As time evolves, under the action of
surface tension, the total liquid-vapor interface length decreases
due to the mergence of small domains, then the discretization errors
decrease. With the increase of precision, variations of total energy
$\Delta e_{tol}(t)$ decreases. We can, therefore, come to the
conclusion that WFFT scheme with higher-order filter has more
advantage in guaranteeing energy conservation than the one with
lower-order filter and other FD schemes used above.

\section{Simulation results and analysis}

In this section, two kinds of typical benchmarks are performed to
validate the physical properties of the thermal multiphase model and
the newly proposed algorithm. The first one is related to a planar
interface. The second one is related to a circular interface.

\subsection{Coexistence curve, spurious velocities and interfacial width}

To check if the thermal LB multiphase model can correctly reproduce
the equilibrium thermodynamics of the system and the numerical
accuracy of the new scheme, a series of simulations about the
liquid-vapor interface at different temperatures were performed.
Unless otherwise stated, the WFFT scheme with $k_{4}$ is used
throughout our simulations.

Simulations were carried out over a $512\times 4$ domain with PBC in both
directions. The initial conditions are set as
\begin{equation}
\left\{
\begin{array}{cc}
(\rho \text{, }u_{x}\text{, }u_{y}\text{, }T)|_{L}=(\rho _{v}\text{, }0.0%
\text{, }0.0\text{, }0.97)\text{,} & x\leq N_{x}/4\text{;} \\
(\rho \text{, }u_{x}\text{, }u_{y}\text{, }T)|_{M}=(\rho _{l}\text{, }0.0%
\text{, }0.0\text{, }0.97)\text{,} & N_{x}/4<x\leq 3N_{x}/4\text{;} \\
(\rho \text{, }u_{x}\text{, }u_{y}\text{, }T)|_{R}=(\rho _{v}\text{, }0.0%
\text{, }0.0\text{, }0.97)\text{,} & 3N_{x}/4<x\text{,}%
\end{array}%
\right.
\end{equation}%
where $\rho _{v}=0.80$ and $\rho _{l}=1.20$ are the theoretical values at $%
T=0.99$. Parameters are set to be $\tau =10^{-2}$, $K=0$ and others
are unchanged. The initial temperature is set to be $0.97$, but
dropping by $0.02$, when the equilibrium state of the system is
achieved. Simulations were then run until the temperature had
reduced to $0.87$.
\begin{figure}[tbp]
\center {
\epsfig{file=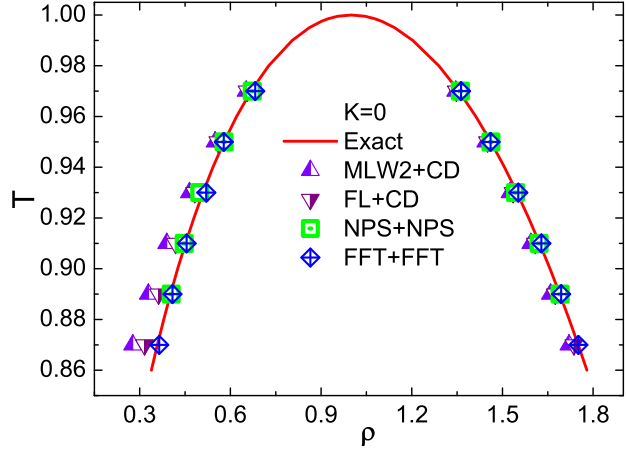,bbllx=2pt,bblly=1pt,bburx=572pt,bbury=418pt,
width=0.5\textwidth,clip=}} \caption{(Color online) Comparisons of
liquid-vapor coexistence curves from
LB simulations and Maxwell construction. Here the surface tension parameter $%
K$ is set to be $0$. }
\end{figure}
\begin{figure}[tbp]
\center {
\epsfig{file=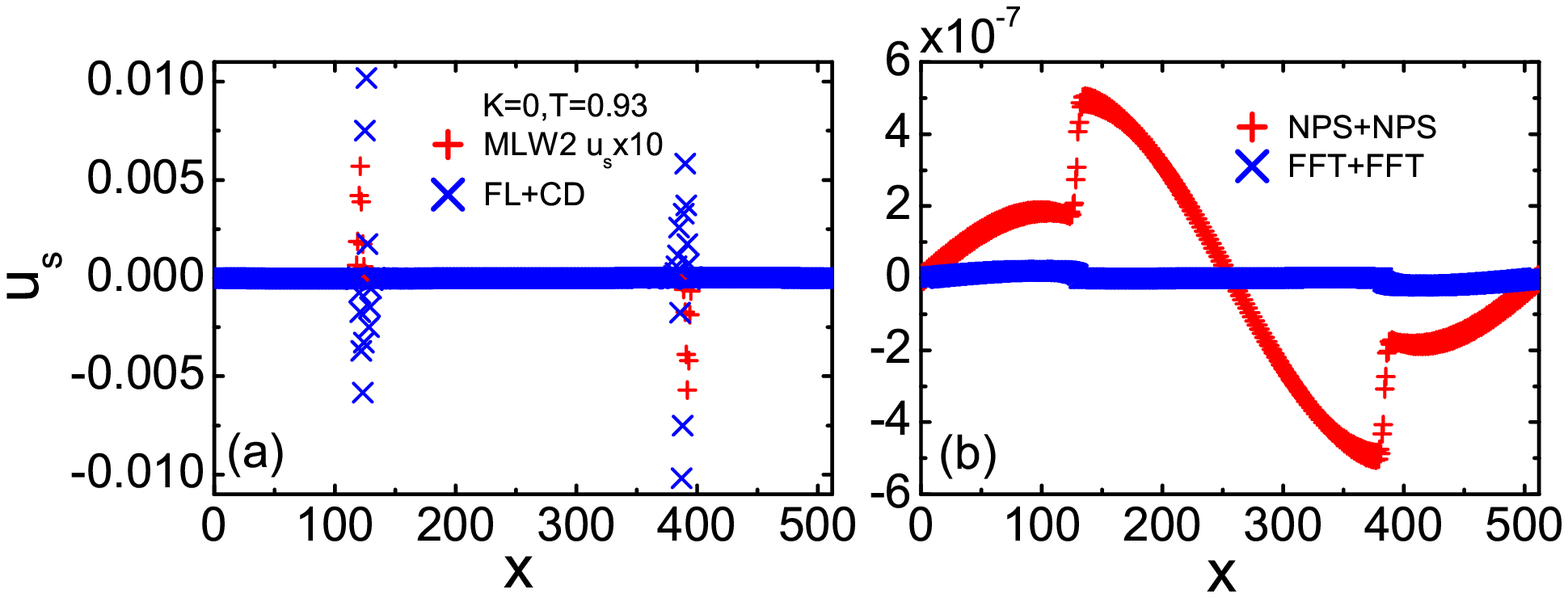,bbllx=5pt,bblly=2pt,bburx=566pt,bbury=217pt,
width=0.7\textwidth,clip=}}
\caption{(Color online) Velocity profiles obtained using various schemes at $%
T=0.93$ with $K=0$. }
\end{figure}
\begin{figure}[tbp]
\center {
\epsfig{file=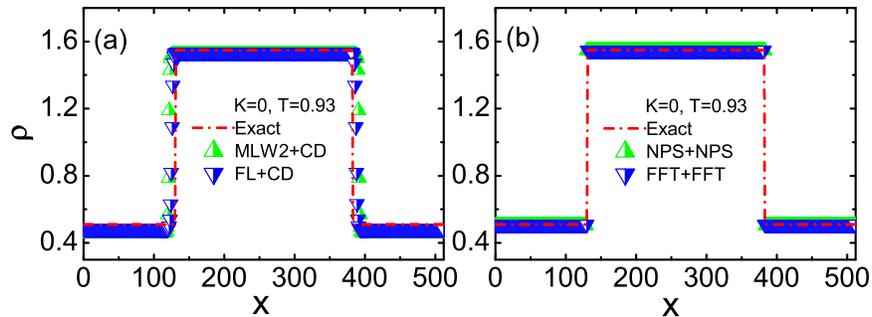,bbllx=5pt,bblly=2pt,bburx=566pt,bbury=217pt,
width=0.70\textwidth,clip=}} \caption{(Color online) Density
profiles obtained using different schemes at $T=0.93$ with $K=0$. }
\end{figure}

In Fig. 9, the liquid-vapor coexistence curves from LB simulations
using various numerical schemes at different temperatures are
compared to the theoretical predictions from Maxwell construction.
One can see that when using the WFFT and the NPS schemes, the
results are closer to the theoretical phase diagram. Nevertheless,
when the temperature is lower than $0.87$, the NPS scheme becomes
unstable. Physically, this is owing to the sharp interface when
$K=0$ (see Fig. 11) and the nature of the VDW EOS, since large
density ratio occurs when the temperature is much lower than the
critical one. From another perspective, it demonstrates that the
WFFT algorithm has a better numerical stability for this test.
Results from the MLW2 and the FL schemes deviate remarkably from the
theoretical values, especially for the vapor branch at lower
temperatures. Besides physical reasons listed above, numerical
accuracy of these two schemes is also an important factor.

\begin{figure}[tbp]
\center {
\epsfig{file=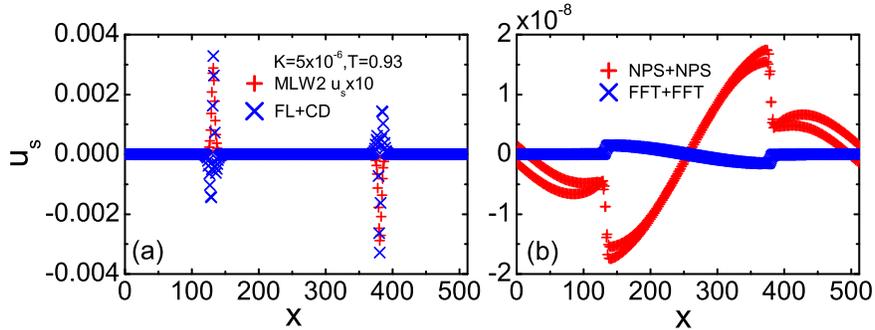,bbllx=5pt,bblly=2pt,bburx=566pt,bbury=217pt,
width=0.7\textwidth,clip=}}
\caption{(Color online) Velocity profiles obtained using various schemes at $%
T=0.93$ with $K=5\times 10^{-6}$. }
\end{figure}

The velocity and density profiles at $T=0.93$ are shown in Fig. 10
and Fig. 11, respectively. As one can see in Fig. 10, for all
schemes, spurious velocities exist and reach their maxima near the
interface regions. However, the maximum spurious velocities obtained
from different schemes are greatly different. For the MLW2 and the
FL schemes, the maxima of $u^{s}$ are on the order of $10^{-4}$ and
$10^{-2}$, respectively. A significant reduction of the $u^{s}$ is
achieved by using the NPS scheme, decreasing the maximum to about
$5\times 10^{-7}$. Through the usage of the WFFT algorithm, $u^{s}$
is further reduced by an order of magnitude compared to NPS scheme.
Density profiles in Fig. 11(a) indicate that spurious interfaces
(scatter symbols near the interfaces) have been produced when using
the MLW2 scheme or the FL scheme because of excess numerical
diffusion, which does not provide us a clear picture of phase
separation, especially when the temperature is close to the critical
value. This feature is not present when using the NPS scheme or the
WFFT scheme (see Fig. 11b).

\begin{figure}[tbp]
\center {
\epsfig{file=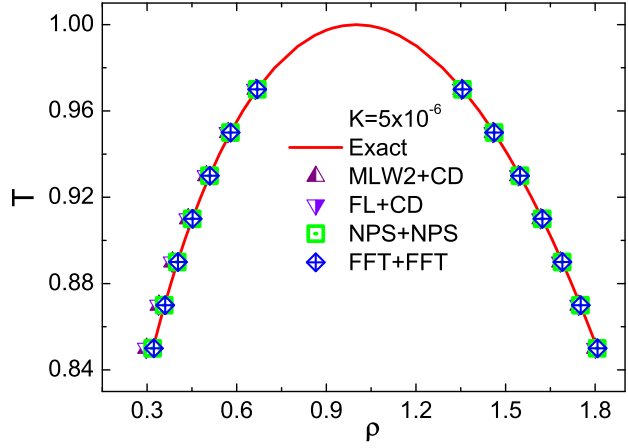,bbllx=4pt,bblly=1pt,bburx=573pt,bbury=413pt,
width=0.5\textwidth,clip=}} \caption{(Color online) Comparisons of
coexistence curves from LB
simulations and Maxwell construction. Here $K$ is set to be $5\times 10^{-6}$%
. }
\end{figure}

For all numerical schemes, the strength of surface tension plays an
important role in reducing spurious velocities, as shown in Fig. 12.
In the case of $K=5\times 10^{-6}$, the amplitudes of $u^{s}$ can be
reduced by a factor of approximately $10$ with respect to the case
of $K=0$. Subsequent
simulations indicate that $u_{\max }^{s}$ will decrease to $10^{-12}$ when $%
K $ increases to $10^{-5}$. This is due to the existence of a wider
interface and a smaller density gradient in the interface region
when $K$ increases. With the decrease of spurious velocities, a more
accurate phase diagram, especially in the vapor branch, is also
achieved (see Fig. 13), even in the MLW2 case and the FL case.
\begin{figure}[tbp]
\center {
\epsfig{file=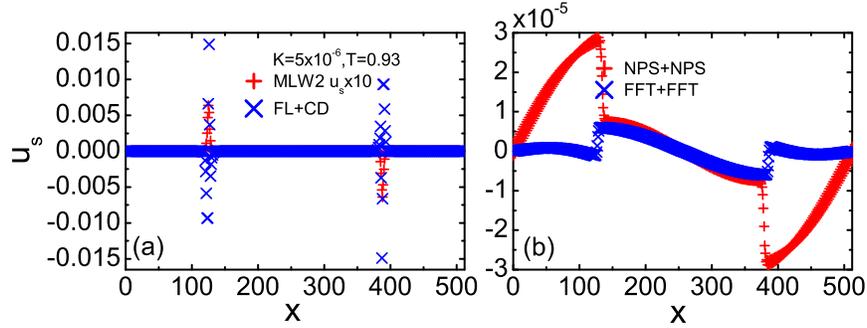,bbllx=5pt,bblly=2pt,bburx=572pt,bbury=220pt,
width=0.70\textwidth,clip=}}
\caption{(Color online) Velocity profiles obtained using various schemes at $%
T=0.85$ with $K=5\times 10^{-6}$. }
\end{figure}
\begin{figure}[tbp]
\center {
\epsfig{file=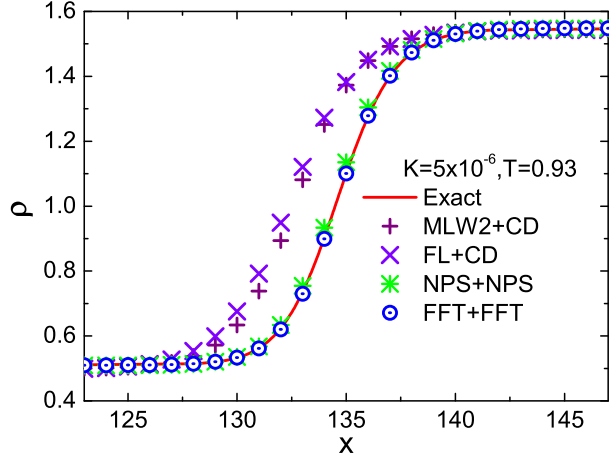,bbllx=1pt,bblly=4pt,bburx=576pt,bbury=433pt,
width=0.5\textwidth,clip=}} \caption{(Color online) Equilibrium
density profiles across liquid-vapor interface at $T=0.93$ with
$K=5\times 10^{-6}$. }
\end{figure}

Besides the strength of surface tension, temperature is another key
factor affecting spurious velocities, as displayed in Fig. 14. For
all numerical schemes, a lower temperature makes larger spurious
velocities. It is worth mentioning that, at the same temperature,
the WFFT algorithm also allows to reduce $u^{s}$ of about an order
of magnitude compared to NPS scheme.

In our simulations, so far, we have not discussed in detail the
width of interface. According to the VDW theory, the interface width
$l$, can be determined by numerically solving the following integral
for a planar interface
\cite{Interface-2-PRA-1975,Interface-3-PRA-1979}
\begin{equation}
l=x-x_{0}=-\frac{1}{\sqrt{2}(a/K)^{1/2}}\int_{\rho ^{\ast }(x_{0}^{\ast
})}^{\rho ^{\ast }(x^{\ast })}\frac{d\rho ^{\ast }}{[\Phi ^{\ast }(\rho
^{\ast })-\Phi ^{\ast }(\rho _{l}^{\ast })]^{1/2}}\text{,}
\end{equation}%
where $\rho ^{\ast }=\rho b$, $T^{\ast }=bT/a$ and,
\begin{equation}
\Phi ^{\ast }=\rho ^{\ast }\xi -\rho ^{\ast }T^{\ast }[\ln (1/\rho ^{\ast
}-1)+1]-\rho ^{\ast 2}\text{,}
\end{equation}%
\begin{equation}
\xi =T^{\ast }\ln (1/\rho _{s}^{\ast }-1)-\rho _{s}^{\ast }T^{\ast }/(1-\rho
_{s}^{\ast })+2\rho _{s}^{\ast }\text{, }s=v\text{, }l
\end{equation}%
with $a=9/8$, $b=1/3$ in this model. Note that the solution of the
above equations gives the exact density profile for a planar
interface for any value of $T$. Equilibrium density profiles across
the liquid-vapor interface at $T=0.93$ from LB simulations versus
results from VDW theory are shown in Fig. 15. It is clear that
although the liquid and the vapor densities calculated from the MLW2
and the FL schemes coincide with the theoretical ones, neither the
MLW2 nor the FL scheme produces the correct interface profile. The
wider interface in these two cases is due to the excess numerical
diffusion, as shown in Fig.11a. The NPS approach leads to a small
deviation from the VDW theory, while the WFFT scheme presents a
perfect consistency with the theoretical solution. In Fig. 16 we
display density
profiles obtained from WFFT algorithm with $K_{1}=5\times 10^{-6}$ and $%
K_{2}=10^{-5}$, respectively. As expected, the interface becomes wider as $K$
increases. A wider interface decreases the density gradient in the interface
region and helps to stabilize the liquid-vapor system at lower temperature.
\begin{figure}[tbp]
\center {
\epsfig{file=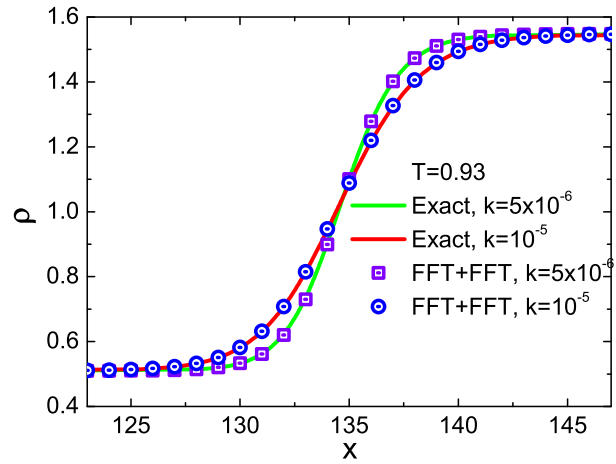,bbllx=1pt,bblly=4pt,bburx=576pt,bbury=433pt,
width=0.5\textwidth,clip=}} \caption{(Color online) Equilibrium
density profiles across the liquid-vapor interface at $T=0.93$ with
$K_{1}=5\times 10^{-6}$ and $K_{2}=10^{-5}$.}
\end{figure}

\subsection{Laplace's law}

\begin{figure}[tbp]
\center {
\epsfig{file=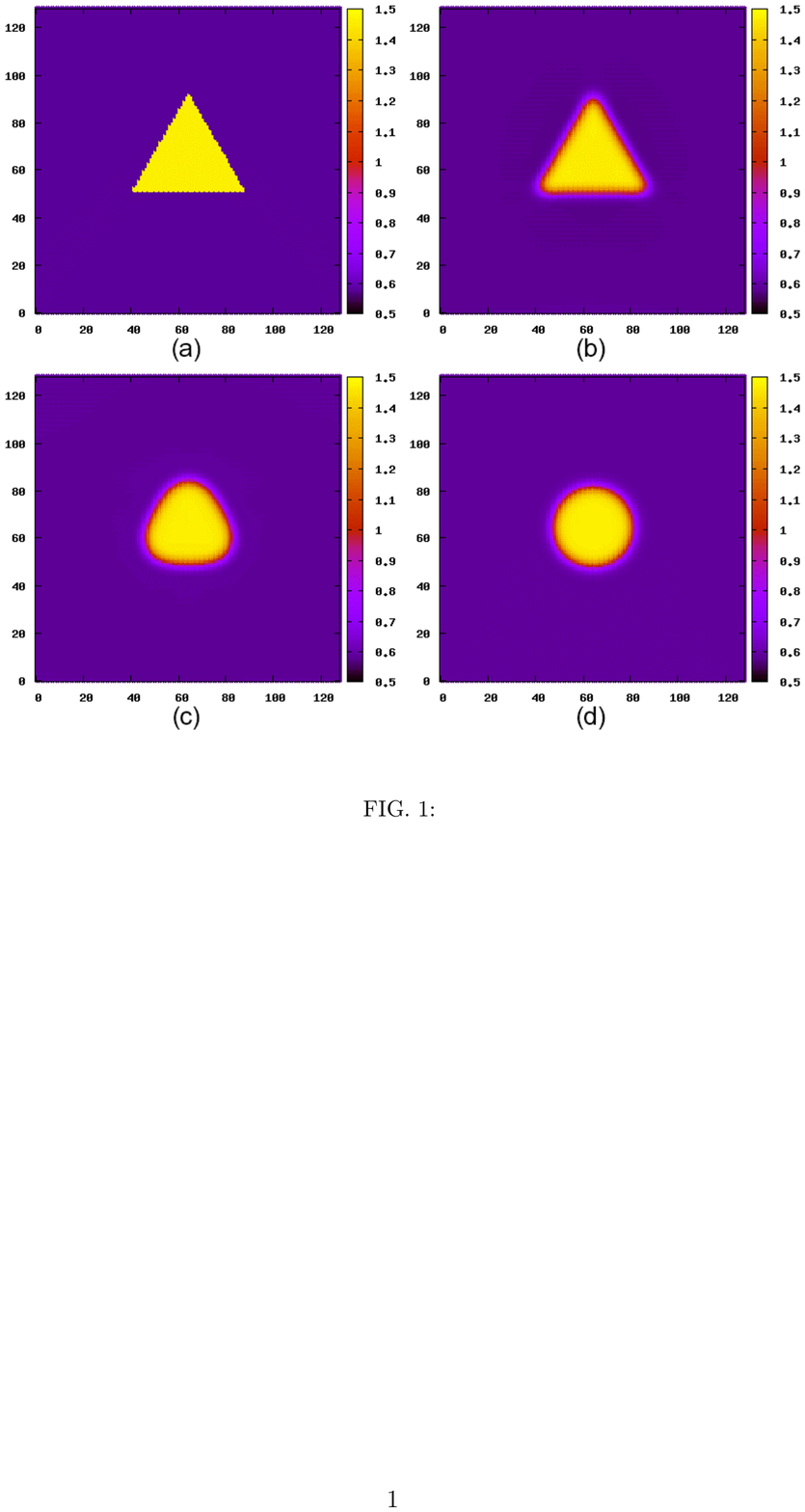,bbllx=119pt,bblly=418pt,bburx=504pt,bbury=772pt,
width=0.80\textwidth,clip=}} \caption{(Color online) Evolution of a
droplet from triangle to circle, where $t=0$ in (a), $t=0.1$ in (b),
$t=1.5$ in (c), and $t=5.0$ in (d). }
\end{figure}

\begin{figure}[tbp]
\center {
\epsfig{file=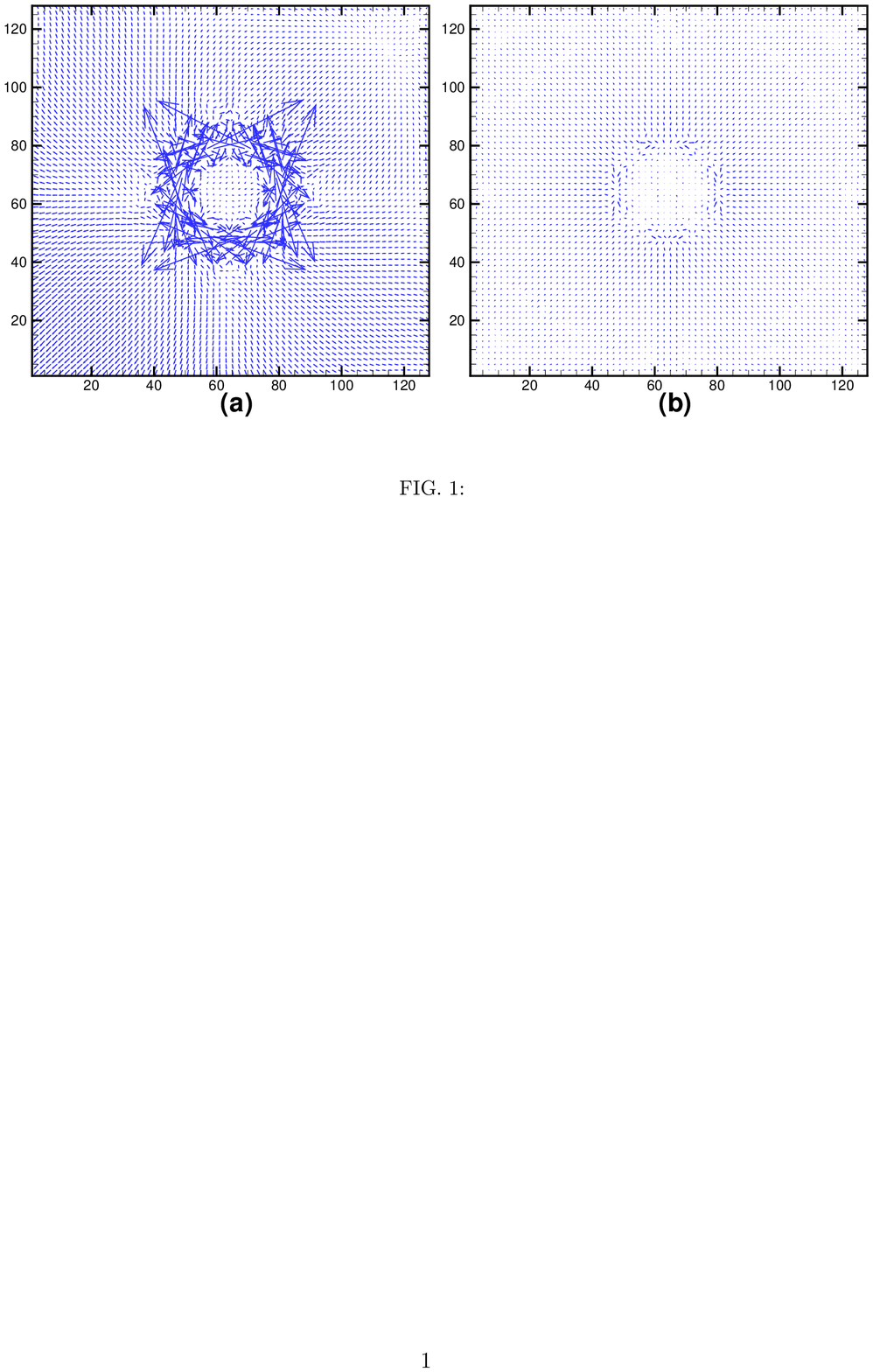,bbllx=79pt,bblly=547pt,bburx=541pt,bbury=771pt,
width=0.8\textwidth,clip=}} \caption{(Color online) Velocity fields
at $t=30.0$ from the NPS scheme in (a) and the WFFT scheme with
$k_{4}$ in (b). }
\end{figure}

In this subsection, we will look at the dynamics of the relaxation of a
deformed droplet driven by surface tension, and investigate the magnitude,
as well as the spatial extent of the spurious currents for the circular
interface. Initially, an equilateral triangular droplet with an initial
sharp interface is placed at the center of the computational domain with $%
N_{x}=N_{y}=128$ lattice units. Initial conditions are given by
\begin{equation}
\left\{
\begin{array}{l}
(\rho \text{, }u_{x}\text{, }u_{y}\text{, }T)|_{in}=(1.46\text{, }0.0\text{,
}0.0\text{, }0.95)\text{,} \\
(\rho \text{, }u_{x}\text{, }u_{y}\text{, }T)|_{out}=(0.58\text{, }0.0\text{%
, }0.0\text{, }0.95)\text{,}%
\end{array}%
\right.
\end{equation}%
where subscripts `in' and `out' indicate macroscopic variables
inside and outside the liquid drop, respectively. PBC are employed
on both the vertical and horizontal boundaries. The surface tension
parameter is $K=10^{-5}$, leaving the others unchanged. After
$3\times 10^{6}$ time steps, the system reaches equilibrium. Contour
plots of the fluid density at four representative times are shown in
Fig. 17. It is clearly seen that due to the effects of surface
tension, the droplet relaxes to a circle slowly.

The velocity fields at $t=30.0$ obtained from the NPS scheme and the
WFFT scheme with $k_{4}$ are plotted in Fig. 18. To illustrate the
structure of the velocity field clearly, the lengths of the velocity
vectors are multiplied by $5\times 10^{5}$. To be seen is that
spurious currents exist in each case and are roughly aligned in the
direction normally to the interface and rapidly disappear away from
the interface. However, the magnitude of the spurious currents are
significantly reduced as the WFFT approach is
used. Figure 19 shows temporal evolution of the maximum velocity $%
u^{s}_{max} $ with the second-order, the fourth-order, the
sixth-order, and the eighth-order FFT schemes, and the NPS scheme.
We can see that, in each case, $u^{s}_{max}$ decreases, and tends to
nearly a constant when $t>20$. More importantly, with the increase
of precision, $u^{s}_{max}$ decreases. There is a decrease of a
factor $22$ for the velocities when using the WFFT scheme with
$k_{4}$ respect to the NPS scheme.

\begin{figure}[tbp]
\center {
\epsfig{file=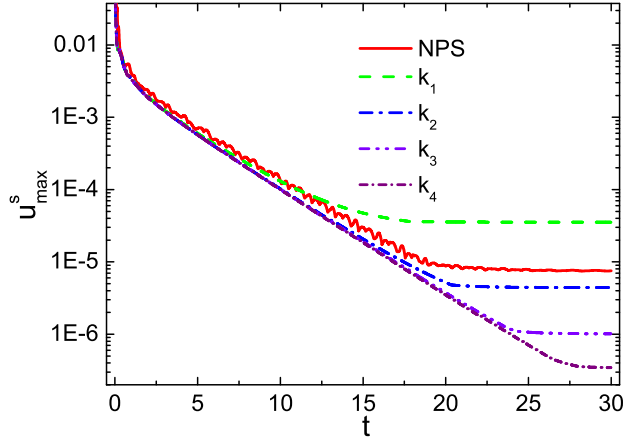,bbllx=5pt,bblly=5pt,bburx=570pt,bbury=470pt,
width=0.5\textwidth,clip=}} \caption{(Color online) Evolution of the
maximum velocity $u^{s}_{max}$
versus time $t$ with the NPS scheme and the WFFT schemes with $k_{1}$, $k_{2}$%
, $k_{3}$, and $k_{4}$.}
\end{figure}

The density and the pressure profiles along the center line of the
droplet are plotted in Fig. 20. In the inner and outer of the droplet, pressures $%
\bar{P}_{in}$ and $\bar{P}_{out}$ are two constants and a rapid change
occurs across the interface. The difference between the two constants is
usually used to compute the surface tension for a given $K$. For this
purpose, we introduce the Laplace law which states
\begin{equation}
\sigma =r\Delta P=r(\bar{P}_{in}-\bar{P}_{out})\text{,}
\end{equation}%
where $\bar{P}_{in}$ is the mean pressure inside the droplet
averaged over all points of $r_{in}<r/2$ from the droplet center and
$\bar{P}_{out}$ is the external mean pressure averaged over all the
points of $r_{out}>3r/2$. In this way, only the particles far from
the interfaces are considered. Surface tension can also be computed
in such a way \cite{kk-1,kk-2,kk-3}
\begin{equation}
\sigma =K\int_{-\infty }^{\infty }(\frac{\partial \rho }{\partial z})^{2}dz%
\text{.}
\end{equation}%
In order to test these relations, a series of simulations with sides ranging
from $48$ to $81$ are run with three different surface tension parameters $%
K_{1}=10^{-5}$, $K_{2}=7.5\times 10^{-6}$ and $K_{3}=5\times 10^{-6}$.
\begin{figure}[tbp]
\center {
\epsfig{file=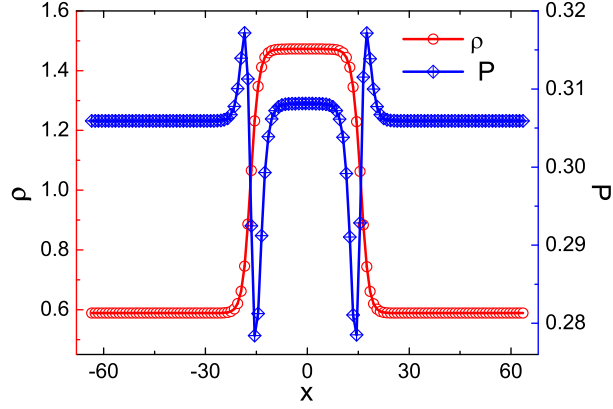,bbllx=10pt,bblly=3pt,bburx=577pt,bbury=380pt,
width=0.5\textwidth,clip=}} \caption{(Color online) Density and
pressure profiles along the center line of a droplet. }
\end{figure}
In Fig. 21, we present a plot of $\Delta P$ versus $1/r$ at $T=0.95$
and linear relation is well satisfied. By measuring the slope, the
surface tension is found to be $\sigma _{1}=1.98\times 10^{-4}$,
$\sigma _{2}=1.77\times 10^{-4}$, and $\sigma _{3}=1.34\times
10^{-4}$, which are in
excellent agreement with the theoretical values, obtained from Eq. (55), $%
\sigma _{1}=1.96\times 10^{-4}$, $\sigma _{2}=1.72\times 10^{-4}$, and $%
\sigma _{3}=1.40\times 10^{-4}$. We mention that the relative error of
surface tension $\varepsilon =\left\vert {\sigma _{LB}-\sigma _{exact}}%
\right\vert /\sigma _{exact}$ increases with the decrease of surface
tension parameter. There are two reasons accounting for this
behavior. Firstly, a larger $K$ will cause a larger surface tension
and a larger pressure difference. This helps to measure the pressure
difference with high accuracy. Secondly, when $K$ decreases larger
spurious velocities will be produced and larger pressure
oscillations will be induced.
\begin{figure}[tbp]
\center {
\epsfig{file=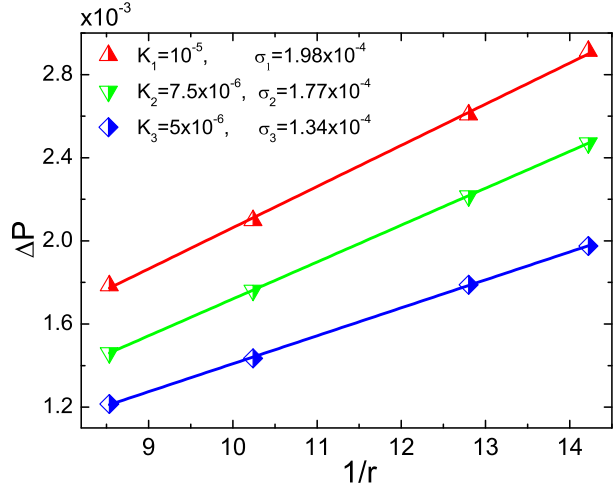,bbllx=2pt,bblly=2pt,bburx=578pt,bbury=455pt,
width=0.5\textwidth,clip=}} \caption{(Color online) Laplace law
tests for three different surface tension parameters. The scattered
symbols are for simulation results and the dashed lines are linear
fits of the scattered symbols. }
\end{figure}

\section{Conclusions and discussions}

In this paper, a thermal LB model for liquid-vapor system is
developed. The present model experienced mainly three stages. It was
originally composed by WT for ideal gas, then developed by GLS by
adding an interparticle force term. Here we propose to use the WFFT
scheme to calculate both the convection term and the external force
term. The usage of the WFFT scheme is detailed and analyzed. It is
found that the lower-order filters, with better numerical stability
and lower accuracy, can effectively reduce the Gibbs phenomenon at
discontinuity, while the higher-order filters help the scheme to
maintain high resolution in smooth regions. One can choose
appropriate filter according to the specific problem. With the
higher-order WFFT algorithm, one can better control the
non-conservation problem of total energy due to spatiotemporal
disterizations. The model has been successfully applied to the
calculation of interfacial properties of liquid-vapor systems. Very
sharp interfaces can be achieved. By adopting the new model the
magnitude of spurious currents can be greatly reduced. As a result,
the phase diagram of the liquid-vapor system obtained from
simulations are more consistent with that from theoretical
calculation. The accuracy of the simulation results is also verified
by the Laplace law.

Besides the numerical effects, both the surface tension and
temperature have also significant influences on the spurious
velocities. A stronger surface tension and/or a higher temperature
can decrease the density gradient near the interfaces and stabilize
the simulations. The analysis presented in this work provides an
convenient way of extending the WFFT approach to multiphase LB
models and to numerical solving partial differential equations. In
further studies we will increase the depth of separation, which the
model can undergo and investigate the similarities and differences
between thermal and isothermal phase separations.

\section*{Acknowledgements}
The authors sincerely and warmly thank the anonymous reviewers for
their valuable comments, encouragements, and suggestions, and we
warmly thank Dr. Victor Sofonea for many instructive discussions. AX
and GZ acknowledge support of the Science Foundations of LCP and
CAEP [under Grant Nos. 2009A0102005, 2009B0101012], National Natural
Science Foundation of China [under Grant No. 11075021]. YG and YL
acknowledge support of National Basic Research Program (973 Program)
[under Grant No. 2007CB815105], National Natural Science Foundation
of China [under Grant No. 11074300], Fundamental research funds for
the central university [under Grant No. 2010YS03], Technology
Support Program of LangFang [under Grant Nos. 2010011029/30/31], and
Science Foundation of NCIAE [under Grant No. 2008-ky-13].

\begin{appendix}

\section{Appendix: numerical accuracies of operators $k_{1}$, $k_{2}$, $%
k_{3} $, and $k_{4}$}

Substituting Eq. (\ref{K-TaylorSeries}) into Eq. (\ref{FFT}), the RHS of Eq. (%
\ref{FFT}) can be expressed as
\begin{eqnarray}
\mathbf{i}k\times \widetilde{f}(k) &=&\frac{\mathbf{i}}{\Delta
x/2}[\sin
(k\Delta x/2)\times \widetilde{f}(k)  \nonumber \\
&&+\frac{1}{6}\sin ^{3}(k\Delta x/2)\times \widetilde{f}(k)  \nonumber \\
&&+\frac{3}{40}\sin ^{5}(k\Delta x/2)\times \widetilde{f}(k)  \nonumber \\
&&+\frac{5}{112}\sin ^{7}(k\Delta x/2)\times \widetilde{f}(k)\cdots
], \label{ikfk}
\end{eqnarray}
Taking IFT of the RHS of the first line of Eq. (\ref{ikfk}) gives
\begin{eqnarray}
\texttt{IFT}[\mathbf{i}k_{1}\times \widetilde{f}(k)] &=&\frac{1}{L}%
\sum_{n=-N/2}^{N/2-1}e^{\mathbf{i}kx_{j}}\times \frac{\mathbf{i}}{\Delta x/2}%
\sin (k\Delta x/2)\times \widetilde{f}(k)  \nonumber \\
&=&\frac{1}{L}\sum_{n=-N/2}^{N/2-1}e^{\mathbf{i}kx_{j}}\frac{e^{\mathbf{i}%
k\Delta x/2}-e^{-\mathbf{i}k\Delta x/2}}{\Delta x}\times
\widetilde{f}(k)
\nonumber \\
&=&\frac{f(x_{j}+\Delta x/2)-f(x_{j}-\Delta x/2)}{\Delta x}  \nonumber \\
&=&f^{\prime }(x_{j})+\frac{1}{24}\Delta x^{2}f^{\prime \prime
\prime }(x_{j})+...
\end{eqnarray}%
It is clear that the FFT scheme with operator $k_{1}$ has a
second-order accuracy in space.

In a similar way, we have
\begin{equation}
\texttt{IFT}[\mathbf{i}k_{2}\times \widetilde{f}(k)]=f^{\prime }(x_{j})+\frac{1}{1920}%
\Delta x^{4}f^{(5)}(x_{j})+...,
\end{equation}%
\begin{equation}
\texttt{IFT}[\mathbf{i}k_{3}\times \widetilde{f}(k)]=f^{\prime }(x_{j})+\frac{1}{%
322560}\Delta x^{6}f^{(7)}(x_{j})+...,
\end{equation}%
\begin{equation}
\texttt{IFT}[\mathbf{i}k_{4}\times \widetilde{f}(k)]=f^{\prime }(x_{j})+\frac{1}{%
92897280}\Delta x^{8}f^{(9)}(x_{j})+...,
\end{equation}%
where $f^{(5)}(x_{j})$, $f^{(7)}(x_{j})$, and $f^{(9)}(x_{j})$
represent the fifth order, the seventh order, and the ninth order
derivatives, respectively. Therefore, the WFFT approach with
$k_{1}$, $k_{2}$, $k_{3}$, and $k_{4}$ has a second-order,
fourth-order, sixth-order, and eighth-order accuracy in space,
respectively.

From another perspective, it should be noted that, the FFT scheme is
not a local scheme or, in other words, $k$ is not a local operator \cite%
{Spectral-Methods-Book-2,Plasma-Book}, since each FFT coefficient is
determined by all the grid point values of $f(x_{j})$, as shown in Eqs.(\ref%
{FT}-\ref{IFT}). Therefore, the FFT scheme is not a finite-point
formula, like the second-order FD is a 3-point formula, or the
fourth order expression, is a 5-point formula; rather, the FFT
scheme is $N$-point formulas. But there are important reasons for
expressing derivatives as local operators. In a continuous space,
the derivative of a function is defined locally. Hence, when
modeling a continuous system with a discrete system, it is desirable
to retain the local character of the derivative. This can be
especially true near boundaries or marked internal inhomogeneities
\cite{Plasma-Book}. From the above derivations, we find that the FFT
scheme with $k_{1}$ corresponds a 3-point FD scheme. Therefore, from
the point of numerical analysis, $k_{1}$, $k_{2}$, $k_{3}$, and
$k_{4}$ can maintain the local characteristic of $k$ in some extent.
Hence, errors arising from the discontinuity are also localized and
the accuracy away from the discontinuity can be ensured.
\end{appendix}

\end{document}